\documentclass[aps,pra,superscriptaddress,twocolumn]{revtex4}

\usepackage{amsmath}
\usepackage{amssymb}
\usepackage{bm} 
\usepackage{color}
\usepackage{hyperref}
\usepackage{graphicx}

\newcommand{\Tr}{\mathrm{\text{Tr}}}
 \newcommand{\brackets}[3]{{}_{#3}\hspace*{-0.2mm}\langle#1|#2\rangle_{#3}}
\newcommand{\ket}[1]{|{#1}\rangle}

\newcommand{\bra}[1]{\langle{#1}|}
\newcommand{\bras}[2]{{}_{#2}\hspace*{-0.2mm}\langle{#1}|}
\newcommand{\kets}[2]{|{#1}\rangle_{#2}\hspace*{-0.2mm}}

\newcommand{\ketbras}[3]{\ket{#1}_{#3}\hspace*{-0.mm}\bra{#2}}

\begin{document}

\title{Discriminating Strength: a bona fide measure of non-classical correlations}

\author{A. Farace}
\affiliation{NEST, Scuola Normale Superiore and Istituto Nanoscienze-CNR, 
 I-56126 Pisa, Italy}   
 
\author{A. De Pasquale}
\affiliation{NEST, Scuola Normale Superiore and Istituto Nanoscienze-CNR, 
 I-56126 Pisa, Italy}     

\author{L. Rigovacca}
\affiliation{Scuola Normale Superiore, 
 I-56126 Pisa, Italy}

\author{V. Giovannetti}
\affiliation{NEST, Scuola Normale Superiore and Istituto Nanoscienze-CNR,  I-56126 Pisa, Italy}

\begin{abstract}
A new measure of non-classical correlations is introduced and characterized. It tests the ability of using a state $\rho$ of a composite system $AB$  as a probe for a {\it quantum illumination} task [e.g. see S. Lloyd, Science {\bf 321}, 1463 (2008)],
in which one is asked to remotely discriminate among the two following scenarios: i) either nothing happens to the probe, or ii) the subsystem $A$ is  transformed via a local unitary $R_A$ whose properties are partially unspecified when producing $\rho$. This new measure can be seen as the discrete version of the recently introduced Intereferometric Power measure [D. Girolami et al.  e-print arXiv:1309.1472 (2013)] and, at least for the case in which $A$ is a qubit, it is shown to coincide (up to an irrelevant scaling factor) with the Local Quantum Uncertainty
measure of D. Girolami, T. Tufarelli, and G. Adesso, 
Phys. Rev. Lett. {\bf 110}, 240402 (2013). Analytical expressions are derived which allow us to formally prove that, within the set of separable configurations, the maximum  value of
our non-classicality measure is achieved over the set of quantum-classical states (i.e. states $\rho$  which admit a statistical unravelling where each element of the associated ensemble is distinguishable via local measures on $B$). 
\end{abstract}
\maketitle

\section{Introduction}

In recent years strong evidences have been collected in support of the fact that 
composite quantum systems 
can exhibit  correlations which, while not being accountable for by  a purely classical   statistical theory, still go beyond  the notion of  quantum entanglement~\cite{MODIREV}.
In the seminal papers by Henderson and Vedral~\cite{vedralzurek}, and Ollivier and Zurek~\cite{OLLZU}, this new form of non-classicality  
was  gauged in terms of a difference of two entropic quantities -- specifically the quantum mutual information~\cite{ENTROPYBOOK} (which accounts for {\it all}  correlations in a bipartite system), and the Shannon mutual information~\cite{COVER}  extractable by performing a generic local measurement on one of the subsystems.
The resulting functional, known  as {\it quantum discord}~\cite{vedralzurek}, enlightens the impossibility of recovering the information contained in a composite quantum system 
by performing local detections only.  It turns out that this intriguing feature of quantum mechanics is not directly related to entanglement~\cite{ENTREV}. Indeed, even though all entangled states are bound to exhibit non-zero
value of quantum discord, examples of  separable (i.e. non entangled) configurations can be easily found  which share the same property
--  zero value of discord identifies only a tiny (zero-measure) subset  of all separable configurations~\cite{FERRA}.
In spite of the enormous effort spent in characterizing this emerging  new aspect of quantum mechanics,   a question which is still open is 
whether and to what extent the new form of quantum correlations identified by quantum discord can be considerd as a {\it resource} and exploited to give some kind of advantage over purely classical means. Due to the variety of contexts where quantum theory proved to be a useful tool for developing new
technological ideas 
(such as information theory, thermodynamics, computation and communication), this gave rise to a number of alternative definitions and quantifiers of discord-like correlations, see e.g.~\cite{MODIREV} and references therein. This proliferation stems also from the difficulty of identifying a measure which is at the same time well defined, easily computable (even for the case of a two-qubit system), and has an operative meaning. As a paradigmatic example, let us recall the geometric discord~\cite{gd} which can be effortlessly computed at the price of being increasing under local operations~\cite{pianiGD}. Some geometric alternatives have been proposed in order to overcome this hindrance. For example one can take the Hilbert-Schmidt distance between the square root of density operators, rather than the density operators themselves~\cite{changGD}, or use different distances such as the trace distance~\cite{Ciccarello2014a} and the Bures distance~\cite{Spehner2013a}. There are also several non-geometric approaches to quantum correlations, both on a fundamental and on an applied level. Among them, let us briefly recall the measurement-induced disturbance~\cite{disturbance} and non-locality~\cite{nonlocality}, which consider the perturbation induced by local von Neumann measurements on non-classically correlated states. On the other hand, the quantum deficit~\cite{Oppenheim2002a} investigates the role of quantum discord in work extraction from a heat bath, while the so-called quantum advantage~\cite{Gu2012a} focuses on quantum discord as the resource allowing quantum communication to be more efficient than classical communication.

Dealing with this complex scenario, here we introduce a new measure of quantum correlations, the \textit{Discriminating Strength} (DS), which turns out to be a valid tradeoff between computability and the fulfillment of the criteria that every good discord quantifier should satisfy~\cite{criteria}. Most importantly, it also possesses a clear operative meaning, being directly connected with the quantum illumination procedures introduced in Refs.~\cite{LLOYD,TAN,JEFF,GU}.
Being the counterpart of the recently introduced Interferometric Power (IP) for continuos variable estimation theory~\cite{blindmetr}, the DS enlightens the benefit gained by quantum state discrimination protocols when general quantum correlations, not necessarily in the form of entanglement, are employed. Finally, 
we provide a formal connection between our new measure and the Local Quantum Uncertainty  Measure  (LQU) introduced in Ref.~\cite{LQU} 
whose operational meaning was not yet
completely understood. Specifically we show that LQU is a special case of DS when the state is used as probe to determine the application of a local unitary which is close to the identity. Furthermore, for qubit-qudit systems 
one can verify that LQU and DS always  coincide up to a proportionality factor. 
The DS, together with the aforementioned IP and LQU, witness a recent burst of attention to the crucial role played by quantum correlations in realm of quantum metrology.

The manuscript is organized as follows. In Sec. \ref{sec:DS} we introduce a paradigmatic state discrimination scheme and we quantify how good a generic state $\rho$ can perform in the discrimination. In Sec. \ref{Sec:Prop} we show that the same quantifier satisfies all the properties required for a bona fide measure of discord. Moreover we present the connection between our measure and the LQU measure and we provide some simple analytical formulas  for some special cases (specifically pure states and qubit-qudits systems). In Sec.~\ref{sec:sep} we focus on the set of separable states and we determine the maximum value of the DS on this set in the qubit-qudits case. Conclusions are left to Sec. \ref{Sec:conc}.

\section{Discriminating Strength} \label{sec:DS}

In order to formally introduce our new measure of non-classicality it is useful to 
  recall the Quantum Chernov Bound (QCB)~\cite{QCB}.
This is an inequality which characterizes the
asymptotic scaling of the minimum error probability~
 $P_{err,min}^{(n)}(\rho_0,\rho_1)$
attainable when discriminating among $n$-copies of two density matrices $\rho_0$ and
$\rho_1$~\cite{QCB}.
By optimizing with respect to all possibile Positive-Operator Valued Measures (POVM) aimed to distinguish among the two possible configurations, 
and assuming a $50\%$ prior probability of getting $\rho_0^{\otimes n}$ or $\rho_1^{\otimes n}$, 
one can write~\cite{HELSTROM} 
 \begin{eqnarray}\label{eq:pminerr}
 P_{err,min}^{(n)}:=\frac{1}{2} (1 - 
\|  \rho_0^{\otimes n} - \rho_1^{\otimes n}\|_1)
\;,
\end{eqnarray}
the optimal detection strategy being the one which discriminates among the negative and non-negative eigenspaces of the operator $\rho_0^{\otimes n} - \rho_1^{\otimes n}$.
For large enough $n$, the dependance of the error probability on the number of copies can be approximated by an exponential 
decay
\begin{equation}
P_{err,min}^{(n)}(\rho_0,\rho_1)\simeq e^{-n\;\xi(\rho_0,\rho_1)} =: Q(\rho_0,\rho_1)^n\;, 
\end{equation}
characterized by the decay constant
\begin{eqnarray}\label{eq:limit}
\xi(\rho_0,\rho_1) : =-  \lim_{n\rightarrow\infty}
\frac{\ln P_{err,min}^{(n)}(\rho_0,\rho_1)}{n}\;.
\end{eqnarray}
Accordingly, the larger is $Q(\rho_0,\rho_1)$
the less distinguishable are the states $\rho_0$ and $\rho_1$. 
The limit in \eqref{eq:limit} corresponds to the QCB bound~\cite{QCB} and reads
 \begin{eqnarray}\label{eq:QCB}
e^{-\xi(\rho_0,\rho_1)} = Q(\rho_0,\rho_1) =  \min_{0 \leq s \leq 1} \Tr \big[ \rho_0^s \rho_1^{1-s} \big]\;, 
  \end{eqnarray} 
which implies  
\begin{eqnarray}\label{bound1}
0 \leq Q(\rho_0,\rho_1) \leq \mbox{Tr}[ \rho_0^{1/2} \rho_1^{1/2}]  \leq 1\;.
\end{eqnarray}
Furthermore if at least one of the two quantum states  $\rho_0$ or $\rho_1$  is pure, then QCB reduces to the Uhlmann's fidelity~\cite{UHL}, i.e. 
  \begin{eqnarray}
  Q(\rho_0,\rho_1)=\mathcal{F}(\rho_0,\rho_1):=\left(\mathrm{Tr}\left[\sqrt{\sqrt{\rho_0} \; \rho_1\; \sqrt{\rho_0}}\right]\right)^2\;.  \label{pure}
  \end{eqnarray}
\begin{figure}[ht]
	\includegraphics[width=0.48\textwidth]{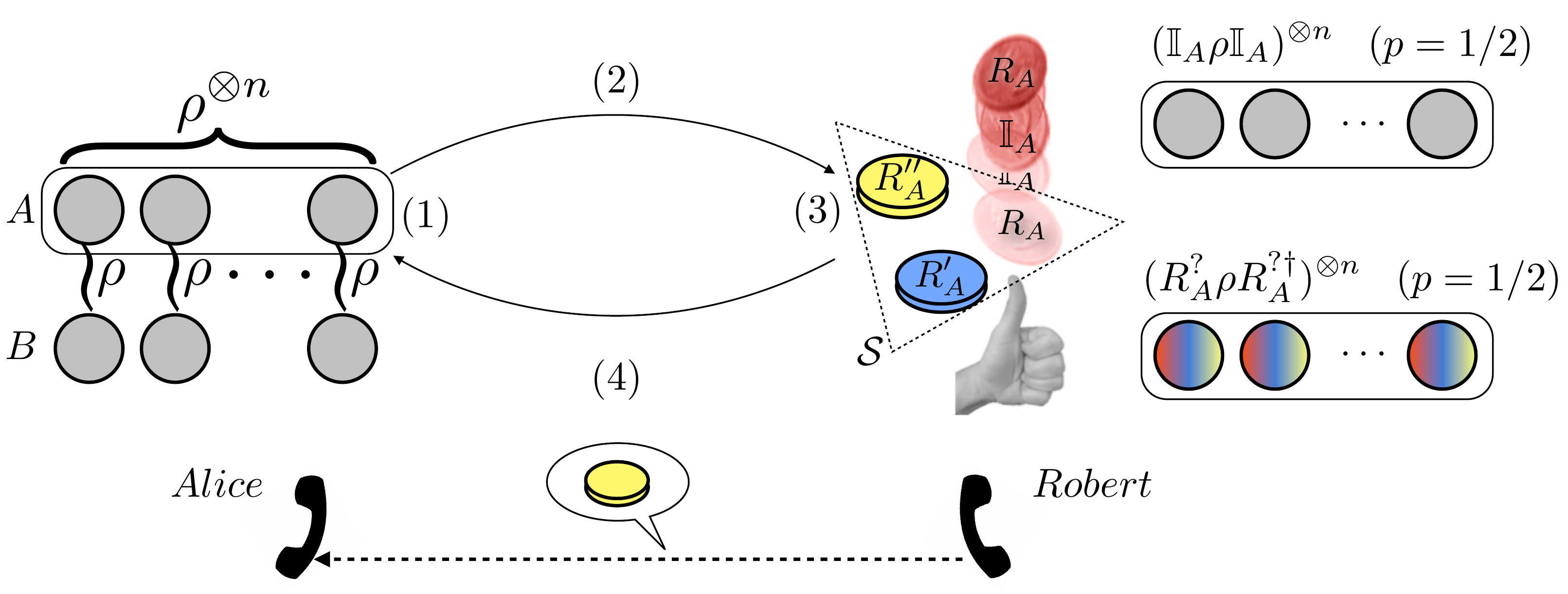}
		\caption{(Color online): sketch of the discrimination problem discussed in the text. 
		(1) A first party (say Alice) prepares  $n$ copies of a bipartite state $\rho$ of a composite system $AB$ and (2) sends the probing subsystems  $A$  to a  second party (say Robert) while keeping the reference subsystems $B$ on her laboratory. 
			(3) Robert can now decide whether or not a certain unitary rotation $R_A$ he has previously selected from a set ${\cal S}$ of allowed transformations,
		 should be  applied (locally) on each one of the probes $A$. (4) After this action the subsystems $A$ are returned to Alice and the chosen $R_A$ is revealed to her. 
		  By exploiting this information and by performing the most general measure on her systems, she has now to determine which option (i.e. the application of $R_A$ or the non application of $R_A$)   Robert has selected. }
		\label{fig:uno}
\end{figure}

Let us now consider the following  {\it quantum illumination} scenario~\cite{LLOYD,TAN,JEFF,GU}. 
A first party (Alice) prepares $n$ copies of a density matrix $\rho$ of a bipartite system  $AB$ composed by a probing component $A$ and a reference component $B$, while
a second party (the non-cooperative target Robert) selects an undisclosed unitary transformation $R_A$ from a set  ${\cal S}$ of allowed transformations.
 Next Alice sends her $n$ subsystems $A$ to Robert who is allowed to do one of the following actions: induce the same rotation $R_A$ 
on each of the $n$ subsystems $A$, or leave them unmodified -- see Fig.~\ref{fig:uno}.  Only after this step  Robert reveals the chosen rotation $R_A$ and sends back  the $A$ subsystems. Alice is now 
requested to guess whether the rotation $R_A$ has been implemented or not, 
 i.e. to \textit{discriminate} between $\rho_0^{\otimes n}=\rho^{\otimes n}$ (no rotation) and
  $\rho_1^{\otimes n} = (R_A \rho R_A^\dag)^{\otimes n}$ (rotation applied). For this purpose of course she is allowed to perform the most general POVM
on the $n$ copies of the transformed states. In particular, as in a conventional interferometric experiment, she might find useful to  exploit the correlations present among the probes $A$ and their corresponding reference counterparts $B$
[it is important to stress however that,  due to the lack of prior info on $R_A$,  Alice cannot perform any  optimization  
with respect to the choice of her initial state  $\rho$]. 
In this scenario we define 
the  ``discriminating strength''  of the state $\rho$ by quantifying Alice's worst possible performance through the quantity 
\begin{equation}
\label{eq:7}
D_{A\rightarrow B}(\rho)  : =  1 - \max_{R_A \in {\cal S} } Q(\rho,R_A \rho R_A^\dagger)\;,
\end{equation}
where the maximization is performed over the set ${\cal S}$ of allowed~$R_A$, and where the symbol $A\rightarrow B$ enlightens the different role played by the two subsystems in the problem -- an asymmetry which is a common trait of the majority of non-classical correlations measures introduced so far~\cite{MODIREV}.

From Eqs.~(\ref{eq:QCB}) and \eqref{eq:7} it is clear that  the higher is $D_{A\rightarrow B}(\rho)$ the better Alice will be able to
determine whether a generic element of ${\cal S}$  has been applied or not to $A$. 
It is a natural guess to expect that the capability shown by the input state $\rho$ of recording the action of an arbitrary local rotation, should increase with the amount of correlations shared between the probe $A$ (which has been affected by the rotation) and the reference $B$ (which has not). This behavior would be analogous to the one displayed by the Interferometric Power measure discussed in~\cite{blindmetr}, which quantifies the worst-case precision in determining the value of a continuous parameter. 
Clearly the choice of ${\cal S}$ plays a fundamental role in our construction: for instance allowing ${\cal S}$ to coincide with the group $\mathbb{U}_A$ of all possible unitary transformations  on $A$, including the identity, would give $D_{A\rightarrow B}(\rho)=0$ for all states $\rho$.
To avoid these pathological results we find it convenient to identify ${\cal S}$ with the special  family of $R_A$ parametrized as
$R_A^\Lambda = \exp[i H_A^\Lambda ]$, 
where 
$H_A^\Lambda$ is a Hamiltonian 
of assigned non-degenerate spectrum represented by the elements of the diagonal matrix 
\begin{eqnarray} \label{defLAMBDA}
\Lambda := \mbox{Diag} \{\lambda_1, \lambda_2, \ldots ,\lambda_{d_A}\}
\;,
\end{eqnarray} 
with $\lambda_1 > \lambda_2 > ... > \lambda_{d_A}$
 ($d_A$ being the dimension of the system $A$) and $\lambda_{1} - \lambda_{d_A} < 2 \pi$ (a condition the latter which can always be enforced by properly relabeling the entries of $\Lambda$).  Accordingly we have 
\begin{eqnarray}  H_A^\Lambda &=& U_A \; 
\Lambda \;  U_A^\dagger\;,\label{diago} \\
 R_A^{\Lambda} &=& U_A \exp[ i \Lambda ] U_A^\dag \;, \label{diago1}
 \end{eqnarray}
where now $U_A$ spans the whole set $\mathbb{U}(d_A)$. For each given choice  of $\Lambda$~(\ref{defLAMBDA})  we thus define the quantity
\begin{equation} \label{defofDS}
D_{A\rightarrow B}^\Lambda(\rho)  : =  1 - \max_{\{ H^\Lambda_A\}} Q(\rho,e^{i H^\Lambda_A} \rho e^{-i H^\Lambda_A})\;,
\end{equation}
the maximization being performed over the set $\{ H_A^\Lambda\}$ of the Hamiltonians of the form~(\ref{diago}). This  measure of discord can be interpreted as an extension to generic non-classical correlations of the entanglement of response, which quantifies the change induced on the state of a composite quantum system by local unitary transformations~\cite{entResp}. In this respect another measure of discord has been recently introduced, the Discord of Response (DR)~\cite{discResp}. The DR is defined in terms of a maximization, over the set of unitary operators endowed with fully non-degenerate spectrum in the roots of the unity, of the Bures distance between the considered state and its evolution under such unitary transformations. Similarly to the DS, the DR accounts for the degree of distinguishability between an assigned quantum state and its evolution under local unitary operators. However, in the case of the DS introduced in this paper, no further limitations, apart from the non-degeneracy, are imposed on the spectrum of the unitary operators.

In the next section we will show that, for all given choices of the spectrum $\Lambda$  the functional~(\ref{defofDS})  fulfills all the requirements necessary for attesting it as a proper measure of non-classical correlations~\cite{MODIREV}. 
\section{Properties}\label{sec:DS as a measure}
\label{Sec:Prop}
In this section we show that the discriminating strength (\ref{defofDS}) is a {\it bona fide} measure of non-classicality. We also clarify the connection between our measure and the
LQU measure introduced by Girolami et al. in Ref.~\cite{LQU}. Finally we 
provide close analytical expressions that, in some special cases, allow one to avoid going through the cumbersome  optimization over the set  $\{ H_A^\Lambda\}$ of the Hamiltonians~(\ref{diago}).

\subsection{DS as a measure of non-classical correlations} \label{subsectionIIIA}
\textbf{Theorem 1:} $D_{A\rightarrow B}^\Lambda(\rho)$ satisfies the following properties:
\begin{enumerate}
\item  it nullifies if and only if $\rho$ is a {\it classical-quantum} (CQ) state~(\ref{CQ})
\begin{eqnarray}\label{CQ}
\rho = \sum_i p_i \; |i\rangle_A \langle i| \otimes \rho_B^{(i)}\;,
\end{eqnarray} 
with $p_i$ being probabilities, $\{|i\rangle_A\}$
being an orthonormal basis of $A$ and $\{ \rho_B(i)\}$ being a collection of density matrices of $B$  (these are the only configurations for which it is possible to recover partial information on the system by measuring $A$, without introducing any perturbation~\cite{MODIREV}); 
\item it is invariant under the action of arbitrary local unitary maps, $W_A$ and $V_B$ on $A$ and $B$ respectively, i.e. 
\begin{equation}
D_{A\rightarrow B}^\Lambda(\rho)=D_{A\rightarrow B}^\Lambda(W_A \otimes V_B \rho W_A^\dagger \otimes V_B^\dagger)\,;
\end{equation}
\item it is non-increasing under any completely positive, trace-preserving (CPT)~\cite{NIELSEN} map $\Phi_B$ on $B$; 
\item it is an entanglement monotone when $\rho$ is pure.
\end{enumerate}

\textit{Proof:} \\

1) $D_{A\rightarrow B}^\Lambda(\rho)=0$ iff there exists at least an element of the set~(\ref{diago})  such that 
$Q(\rho,R_A^{\Lambda}  \rho R_A^{\Lambda\dag} )=1$.
The latter condition is satisfied iff~\cite{QCB} $\rho=R_A^{\Lambda}  \rho R_A^{\Lambda \dag} $.
Being $R_A^{\Lambda}$ endowed with a non-degenerate spectrum, this is equivalent to stating that $\rho$ and $H_A^\Lambda$ are diagonal in the same basis $\{|i\rangle_{A}\}$ of $\mathcal{H}_A$, and thus $\rho$ reduces to a CQ state of the form~(\ref{CQ}).

2)
First note that for every unitary operator $U$ it holds $(U \rho U^\dagger)^s = U \rho^s U^\dagger$. Then, due to the cyclic property of the trace, $V_B$ cancels out with $V_B^\dagger$ in the computation of $Q$. Finally $W_A^\dagger  H_A^\Lambda W_A$ has the same spectrum of $H^\Lambda_A$ so that the maximization domain in \eqref{defofDS} remains unchanged along with the maximum value.

3) This follows from the very definition of the QCB. Indeed, the minimum error probability in \eqref{eq:pminerr} is achieved by optimizing over all possible POVM measurements on $(AB)^{\otimes n}$. Any local map  $\Phi_B$ on $B$ commutes with the phase transformation determined by $H_A^\Lambda$, and thus can be reabsorbed in the measurement process. This modified measurement is at most as good as the optimal one, implying that the asymptotic error probability, and hence $Q$, cannot decrease. This gives 
$D_{A\rightarrow B}^\Lambda(\Phi_B[\rho])\leq D_{A\rightarrow B}^\Lambda(\rho)$.

4) We will prove that if a pure state $|\psi\rangle$ is transformed into another pure state $|\phi\rangle$ by LOCC (Local Operations and Classical Communication), then $D_{A\rightarrow B}^\Lambda(|\phi\rangle)\leq D_{A\rightarrow B}^\Lambda(|\psi\rangle)$. We remind that, due to the purity of the input and output states,  a generic  LOCC transformation which maps the vector 
$|\psi\rangle$ in $|\phi\rangle$
can always be
realized via a single POVM on $A$ followed by a  unitary rotation on $B$ conditioned
by the measurement outcome, see e.g.~\cite{NIELSEN}.
In other words,  we can write
\begin{eqnarray}\label{eq:phi_psi}
|\phi\rangle \langle \phi|=\sum_{j} ({M_j}_A  {V_j}_B)|\psi\rangle \langle \psi|({M_j}^\dagger_A  {V_j}^\dagger_B)\;,
\end{eqnarray}
where $\{{M_j}_A\}$ is a set of Kraus operators on $A$ ($\sum_{j}{M_j}^\dagger_A {M_j}_A=\mathbb{I}_A$),  
and $\{{V_j}_B\}$ is a set of unitary operators on $B$.
Introducing  the set of probabilities  $\{p_j\}=\{\brackets{\psi|{M_j}^\dagger_A {M_j}_A}{\psi}{}\}$, from \eqref{eq:phi_psi}
it follows
that for all $j$ corresponding to $p_j \neq 0$ we must have 
\begin{eqnarray} \label{NEWEQ}
{M_j}_A   {V_j}_B|\psi\rangle  = \sqrt{p_j}|\phi\rangle \,\quad \forall \, j \; \, \mbox{s.t.}  \; \,  p_j \neq 0\;.
\end{eqnarray}
Observe also that for each $H_A^\Lambda$, there exists an $ H_B^\Lambda$ which has the same components in the Schmidt basis of $|\psi\rangle$, that is
\begin{equation}
\brackets{\psi|e^{i H_A^\Lambda  } \otimes\mathbb{I}_B}{\psi}{} = \brackets{\psi| \mathbb{I}_A \otimes e^{i  H_B^\Lambda } }{\psi}{}\,.
\end{equation}
From Eq.~(\ref{pure}) it follows then that  for pure input states the maximization over all $H_A^\Lambda$ is equivalent to a maximization over all $H_B^\Lambda$. This allows to write
\begin{eqnarray}\label{eq:pure_ds}
D_{A\rightarrow B}^\Lambda(|\psi\rangle_{})
&=& 1 - \max_{\{H_A^\Lambda\}}\big| \brackets{\psi| e^{i H_A^\Lambda  }}{\psi}{}\big|^2  \\
&=&1 - \max_{\{H_B^\Lambda\}}\big| \brackets{\psi| e^{i H_B^\Lambda }}{\psi}{}\big|^2 \nonumber\\
&=&1 - \big| \brackets{\psi| e^{i \tilde H_B^\Lambda }}{\psi}{}\big|^2, \label{NEWEQ3}
\end{eqnarray}
where we $\tilde H_B^\Lambda$ labels the Hamiltonian for which the maximum is reached.
Along the same lines, we have 
\begin{eqnarray}
 D_{A\rightarrow B}^\Lambda (|\phi\rangle_{})
&=&
 1 - \max_{\{H_B^\Lambda\}} \big| \brackets{\phi|   e^{i H_B^\Lambda}}{\phi}{} \big|^2\\  \nonumber 
 &=&1- 
 \sum_j \frac{1}{p_j} \max_{\{H_B^\Lambda\}} \big| \brackets{\psi| {M_j}_A^\dagger {M_j}_A  e^{i  H_B^\Lambda} }{\psi}{} \big|^2 \;,
 \end{eqnarray} 
where the second identity follows from
 Eq.~(\ref{NEWEQ}) by  absorbing 
the  unitary operator $V_{jB}$ into the maximization
over $H_B^\Lambda$. The rhs of the latter expression can be bounded from above by noticing that 
the maximum of a given function is greater than the function evaluated at a given point.
In particular we have 
\begin{eqnarray}
D_{A\rightarrow B}^\Lambda (|\phi\rangle ) 
\leq 1- \sum_i \frac{1}{p_j} \big| \brackets{\psi| {M_j}_A^\dagger {M_j}_A e^{i \tilde H_B^\Lambda}}{\psi}{} \big|^2 \;,
\end{eqnarray}
where $\tilde H_B^\Lambda$ has been introduced in Eq.~(\ref{NEWEQ3}). Finally, applying the Cauchy-Schwarz inequality we get 
\begin{eqnarray}
D_{A\rightarrow B}^\Lambda (|\phi\rangle )  &\leq&  1- \big| \brackets{\psi| \sum_j {M_j}_A^\dagger {M_j}_A e^{i \tilde H_B^\Lambda} }{\psi}{ } \big|^2 
\\ \nonumber 
&=& 1- \big| \brackets{\psi|e^{i \tilde H_B^\Lambda} }{\psi}{} \big|^2 = D_{A\rightarrow B}^\Lambda (|\psi\rangle),
\end{eqnarray}
hence concluding the proof.  $\blacksquare$

\subsection{A formal connection between  DS and LQU  measures} \label{sec:conn}
The LQU measure of non-classical correlations was introduced in Ref.~\cite{LQU}. Given a state $\rho$ of the bipartite system $AB$ it can be computed  as
\begin{eqnarray} \label{FF}
{\cal U}^{\Lambda}_{A\rightarrow B} (\rho) &=& \min_{\{H^\Lambda_A\}}  {\cal I}(\rho, H^\Lambda_A) \;, 
\end{eqnarray} 
where 
\begin{eqnarray} \label{WYAN}
{\cal I}(\rho, H^\Lambda_A) &:=&
 \mbox{Tr}[H_A^\Lambda  \rho H_A^\Lambda -  \sqrt{\rho} H_A^\Lambda \sqrt{\rho} H_A^\Lambda  ]\;,
\end{eqnarray} 
is the Wigner-Yanase skew information~\cite{WYS} and where, as in Eq.~(\ref{defofDS}), the maximum is taken over  the set $\{ H_A^\Lambda\}$ of the Hamiltonians~(\ref{diago}). 
A connection between~(\ref{FF}) and our DS measure follows by  taking a formal expansion  of Eq.~(\ref{defofDS}) with respect to $\Lambda$, i.e.
\begin{eqnarray} 
&&D_{A\rightarrow B}^\Lambda(\rho)  =  1 - \max_{\{ H_A^\Lambda\}}  \min_{0 \leq s \leq 1} \Tr \big[ \rho^s e^{i  H_A^\Lambda} \rho^{1-s} e^{-i H_A^\Lambda} \big]\nonumber \\
&&=  - \max_{\{H_A^\Lambda\}}   \min_{0 \leq s \leq 1} \Tr \big[ \rho^s H_A^\Lambda \rho^{1-s} H_A^\Lambda- H_A^\Lambda  \rho  H_A^\Lambda  \big] +O(\Lambda^3)\nonumber \\
&&=  - \max_{\{H_A^\Lambda\}}   \Tr \big[ \sqrt{\rho} H_A^\Lambda \sqrt{\rho} H_A^\Lambda- H_A^\Lambda  \rho  H_A^\Lambda  \big] +O(\Lambda^3) \nonumber \\
&&=  \min_{\{H_A^\Lambda\}}   \Tr \big[ H_A^\Lambda  \rho  H_A^\Lambda - \sqrt{\rho} H_A^\Lambda \sqrt{\rho} H_A^\Lambda \big] +O(\Lambda^3) 
\nonumber \\
&&\qquad \qquad \; =\;\; {\cal U}^{\Lambda}_{A\rightarrow B} (\rho)  +O(\Lambda^3) \;, \label{formalcon}
 \end{eqnarray} 
 where in the third identity we used the following property.\\

 {\bf Lemma 1:}
 {\it Given $\rho$ a density matrix and $\Theta=\Theta^\dag$ a Hermitian operator we have }
 \begin{eqnarray} 
&&\min_{0 \leq s \leq 1} \Tr \big[  \rho^s \Theta \rho^{1-s} \Theta \big]   = \Tr \big[  \rho^{1/2} \Theta \rho^{1/2} \Theta \big] \;.
  \end{eqnarray} 
\\
\textit{Proof:} Expressing $\rho$ in terms of its eigenvectors $\{|\psi_\ell\rangle\}$  we can write
\begin{eqnarray}
&& \min_{0 \leq s \leq 1} \Tr \big[  \rho^s \Theta \rho^{1-s} \Theta \big] = \sum_{\ell} c_\ell |\langle \psi_\ell| \Theta | \psi_\ell\rangle|^2  \nonumber  \\
&&\qquad \qquad\quad +\min_{0 \leq s \leq 1} \sum_{\ell<\ell'} (c_\ell^s c_{\ell'}^{1-s} +c_{\ell'}^s c_{\ell}^{1-s}  ) \nonumber 
 |\langle \psi_\ell| \Theta | \psi_{\ell'}\rangle|^2\;,
\end{eqnarray} 
where $\{c_\ell\}$ are the eigenvalues of $\rho$ which have being organized  in decreasing order (i.e. $c_\ell \geq c_{\ell'}$ for $\ell\leq \ell'$). 
The thesis then follows by simply noticing that for all couples  $\ell<\ell'$, the functions $f(s) = c_\ell^s c_{\ell'}^{1-s} +c_{\ell'}^s c_{\ell}^{1-s}$ reach their minima 
for $s=1/2$ (indeed their first derivative $f'(s) = (c_\ell^s c_{\ell'}^{1-s} - c_{\ell'}^s c_{\ell}^{1-s}) \ln(c_\ell/c_{\ell'})$ are non-negative for $s\geq 1/2$ and non-positive for $s\leq 1/2$). $\blacksquare$
\\ 

Equation~(\ref{formalcon}) establishes a formal connection between our DS measure and the LQU measure, providing hence a clear operational interpretation for the latter.
Specifically  the LQU can be seen as the DS measure of a discrimination process where $\Lambda$ is a small quantity, i.e. where the allowed
rotations  $R_A^{\Lambda}$ of Eq.~(\ref{diago1})  are small perturbation of the identity operator. 
As we shall see in Sec.~\ref{sec:DS_qubit_qudit}, the relation among DS and LQU becomes even more stringent when $A$ is a qubit system: indeed, in this special case,
independently from the dimensionality of $B$, the two measure are proportional.

\subsection{Dependence upon $\Lambda$}
According to Sec.~\ref{subsectionIIIA} all choices of the matrices $\Lambda$ as in Eq.~(\ref{defLAMBDA}) provide a proper measure of non-classicality for the states $\rho$. 
Even though one is tempted to conjecture that the case where  $\Lambda$ has  an harmonic spectrum (i.e. $\lambda_k-\lambda_{k-1} = \mbox{const}$ for all $k=2, 3, \cdots, d_A$)
should be somehow optimal (i.e. yield  a more accurate measure of non-correlations), 
the relations among these different DSs  at present are not clear and indeed it might be possible that no absolute ordering can be established among them (this is 
very much similar to what happens for the LQU of Ref.~\cite{LQU}).
Here we simply notice that since QCB is invariant  under constant shifts in the local Hamiltonian spectrum, i.e. 
$Q(\rho,e^{i H^\Lambda_A} \rho e^{-i H^\Lambda_A})=Q(\rho,e^{i (H^\Lambda_A + b \mathbb{I}_A)} \rho e^{-i (H^\Lambda_A + b \mathbb{I}_A)})$, 
for all incoming states $\rho$ and for $b\in \mathbb{Re}$, we can always  add a constant to $\Lambda$ at convenience without 
affecting the corresponding DS measure, i.e. 
\begin{equation} \label{defofDS111}
D_{A\rightarrow B}^\Lambda(\rho)  = D_{A\rightarrow B}^{\Lambda+b}(\rho) \;, \quad \forall \rho\;. 
\end{equation}

\subsection{Discriminating strength for pure states}

Let $\ket{\psi}$ be a pure state of $AB$ with Schmidt decomposition~\cite{NIELSEN}  given by
\begin{equation}\label{eq:Schmidt}
\ket{\psi}=\sum_{j=1}^{\min\{d_A, d_B\}} \sqrt{q_j}\ket{j}_A\ket{j}_B\,,
\end{equation}
being $\{ |j\rangle_A\}$ and $\{ |j\rangle_B\}$ orthonormal
sets of $A$ and $B$, respectively ($d_{A,B}$ being the dimensionality of $A,B$). From Eq.~(\ref{eq:pure_ds}) it follows that in this case 
the discriminating strength can be written as
 \begin{eqnarray}\label{eq:Dpure1}
D_{A^\rightarrow B}^{\Lambda}(\ket{\psi})&=&1-\max_{\{H_A^\Lambda\}} \left|\sum_{j}q_j\bras{j}{A}e^{i H_A^\Lambda}\kets{j}{A}\right|^2\nonumber\\&=&1-\max_{\{H_A^\Lambda\}} \left|\Tr[\rho_A e^{i H_A^\Lambda}]\right|^2\;,
\end{eqnarray} 
where $\rho_A=\Tr_B[\ketbras{\psi}{\psi}{A}]$ is the reduced state of $|{\psi}\rangle$ on $\mathcal{H}_A$.
From the spectral decomposition~(\ref{diago}) of $H_A^\Lambda$, 
one can perform the trace in~\eqref{eq:Dpure1} over the eigenbasis  of $\Lambda$ 
and get
 \begin{eqnarray} \label{eqnew1}
D_{A\rightarrow B}^\Lambda (\ket{\psi}) =
1-\max_{\{M\}} \left|\sum_{k}\left(\sum_{j}M^{(k|j)} q_j\right) e^{i \lambda_k}\right|^2,
\end{eqnarray} 
where now the maximization is performed over the set of the double stochastic matrices $M$ with elements $M^{(k|j)}=\bras{\lambda_k}{A}U_A^\dagger\kets{j}{A}\bra{j}U_A\kets{\lambda_k}{A}$. We remind that according to the Birkhoffs theorem~\cite{Birkhoff} $M$ can be written 
as a convex combination of permutation matrices $\Pi_\alpha$ (corresponding to the permutation $\pi_\alpha$), i.e. 
\begin{equation}
B = \sum_{\alpha} p_\alpha \Pi_\alpha \quad \mbox{with} \quad \sum_\alpha p_\alpha=1\,.
\end{equation}
Therefore, we can rewrite Eq.~(\ref{eqnew1}) as 
\begin{eqnarray}
\label{eq:permutations}
D_{A\rightarrow B}^\Lambda(\ket{\psi})&=&1-\max_{\{p_\alpha\},\{\Pi_\alpha\} } \left|\sum_{\alpha,k }p_\alpha \sum_{j} \Pi_\alpha^{(k|j)}q_{j} \; e^{i \lambda_k }\right|^2 \nonumber \\
&=&1-\max_{\{p_\alpha\},\{\pi_\alpha\} } \left|\sum_{\alpha}p_\alpha \sum_{k} q_{\pi_\alpha[k]}  e^{i \lambda_k }\right|^2\,. \nonumber\\
\end{eqnarray}
Note that if $d_B < d_A$, the number of Schmidt coefficients is smaller than the number of eigenvalues $\lambda_k$. In this case, the expressions above hold as long as one considers the state \eqref{eq:Schmidt} as having $d_A - d_B$ Schmidt coefficients equal to zero, i.e. one must apply the permutations to the set $\{ q_1, ..., q_{d_B}, q_{d_B+1}=0, ..., q_{d_A}=0 \}$. 

By convexity it derives that the optimization over 
the set $\{p_\alpha\}$ in \eqref{eq:permutations} can be explicitly carried out by 
choosing probability sets $\{p_\alpha\}$ which have only a single element greater than zero (and thus equal to $1$), from which we finally derive 
\begin{equation} \label{NEWNEW1}
D_{A\rightarrow B}^\Lambda (|\psi\rangle )=1-\max_{\pi_\alpha} \left|\sum_{k}q_{\pi_\alpha[k]}e^{i\lambda_{k}}\right|^2\;,
\end{equation}
where the maximization over the infinite set of Hamiltonians $H_A^\Lambda$ required by its definition (see Eq.~\eqref{defofDS}) has been replaced by  a maximization over the group of permutations $\{\pi_\alpha\}$ on the set of the Schmidt coefficients $q_j$.

\subsubsection{Hamiltonians with harmonic spectrum} 
If the spectrum of the Hamiltonian $H_A^\Lambda$ is harmonic with fundamental frequency $\omega=|\lambda_{i}-\lambda_{i+1}|   \leq 2\pi/d_A$, Eq.~\eqref{NEWNEW1} can further simplified. More precisely, let us relabel the set of eigenvalues $\{\lambda_i\}$ as
\begin{eqnarray}
&\left\{\lambda_{1-[(d_A+1)/2]}, \lambda_{2-[(d_A+1)/2]}, \ldots,  \lambda_{d_A-[(d_A+1)/2]} \right \}=\nonumber\\\nonumber\\ &\left \{\left(1\!-\! \left[\frac{d_A+1}{2}\right]\right) \! \omega, \!\left(2 \!-\! \left[\frac{d_A+1}{2}\right]\right)\! \omega, \ldots, \!\left(d_A \!-\! \left[\frac{d_A+1}{2}\right]\right)\! \omega \right \}, \quad\;
\end{eqnarray} 
where $[x]$ stands for the integer part of the real parameter $x$.
Let us also reorder the Schmidt coefficients of $\ket{\psi}$ as $q_1 \geq q_2 \geq \ldots \geq q_{d_A}$ (where again some of them must be set to zero if $d_B < d_A$). By representing the phases $e^{i \lambda_k}$ as unitary vectors in the complex space, one derives that the permutation $\pi$ maximizing the sum
in \eqref{NEWNEW1}
is the one which associates $q_1$ to $\lambda_0=0$,  $q_2$ to $\lambda_{1}=\omega$, $q_3$ to $\lambda_{-1}=-\omega$,  $q_4$ to $\lambda_2=2\omega$, $q_5$ to $\lambda_{-2}=-2\omega$, etc., yielding 
\begin{eqnarray}
&&D_{A\rightarrow B}^\Lambda(\ket{\psi}) \\ \nonumber 
&&= 1-\left| \sum_{n=0}^{[(d_A+1)/2]-1} \!\!\!\!\!\!\!\!\!\! q_{2n+1}  e^{-i n \omega}+ \!\!\!\!\! \sum_{n=1}^{d_A-[(d_A+1)/2]} \!\!\!\!\!\!\!\!\!\! q_{2n} e^{i n \omega}\right|^2\,.
\end{eqnarray}

\subsection{Discriminating strength for qubit-qudit systems}\label{sec:DS_qubit_qudit}
We conclude the Section by considering the case in which subsystem $A$ is  given by a single qubit, and determine a closed expression for the discriminating strength.

Exploiting the gauge invariance~(\ref{defofDS111}) we set, 
without loss of generality,  $\Lambda=\mbox{Diag}\{-\lambda,\lambda\}$ and parameterize the set of local Hamiltonians acting on $A$  as
$H_A^\Lambda  = \lambda\;  \hat n \cdot \vec \sigma_A$, where
$\hat n$ is a unit vector in the Bloch sphere and $\vec{\sigma}_A=(\sigma_{A,1},\sigma_{A,2}, \sigma_{A,3})$ is the vector formed by the Pauli operators. 
 In what follows we will set $\sigma_A^{(\hat{n})}=\hat n \cdot \vec \sigma_A$. 
Under these hypothesis, the QCB can be written as 
\begin{eqnarray}\label{eq:QCB_qubitTrace}
Q(\rho_0,\rho_1) &=&  \min_{s\in[0,1]}
\Tr \big[  \rho^s  e^{i \lambda  \sigma_A^{(\hat{n})} } \rho^{1-s} e^{-i \lambda   \sigma_A^{(\hat{n})}} \big]\nonumber \\
& =& \cos^2\lambda + \min_{s\in[0,1]} \mbox{Tr} [\rho^{s}  \sigma_A^{(\hat{n})}  \rho^{1-s} \sigma_A^{(\hat{n})}]  \; \sin^2\lambda \nonumber \\
& =& \cos^2\lambda +  \mbox{Tr} [\rho^{1/2} \sigma_A^{(\hat{n})}  \rho^{1/2} \sigma_A^{(\hat{n})}]  \; \sin^2\lambda\nonumber \;,
\end{eqnarray}
where in the last passage we  have used  the fact that  $\sigma_A^{(\hat{n})}$ is Hermitian and Lemma 1 to conclude that the minimization in $s$ is solved for $s=1/2$ 
 (see also Ref.~\cite{QCBqubitAndGauss}, footnote 5 on page 11). 
Replacing this into Eq.~(\ref{defofDS}) we finally obtain 
\begin{eqnarray}
D_{A\rightarrow B}^\Lambda(\rho)  &=& \max_{\hat{n}}  \big( 1 -  \; \mbox{Tr} [\rho^{1/2} 
\sigma_A^{(\hat{n})}  \rho^{1/2} \sigma_A^{(\hat{n})}] \big) \;  \sin^2 \lambda \nonumber \\  
 &=& {\cal U}_{A\rightarrow B}^\Lambda (\rho)\;  \frac{\sin^2 \lambda}{\lambda^2} \;, \label{eq:DS_LQU}\end{eqnarray} 
where 
\begin{eqnarray} 
{\cal U}_{A\rightarrow B}^\Lambda(\rho) = 
\lambda^2 \max_{\hat{n}}  \big( 1 -  \; \mbox{Tr} [\rho^{1/2} 
\sigma_A^{(\hat{n})}  \rho^{1/2} \sigma_A^{(\hat{n})}] \big) \;,
\end{eqnarray}
is the LQU measure for a qubit-qudit system~\cite{LQU} -- see Eqs.~(\ref{FF}) and (\ref{WYAN}). The identity~(\ref{eq:DS_LQU}) strengthens the formal connection between DS and LQU detailed in Sec.~\ref{sec:conn} and
provides a simple way to compute the DS for qubit-qudit systems. Indeed using the results of Ref.~\cite{LQU} it follows that 
\begin{eqnarray}\label{EQNUOVA}
D_{A\rightarrow B}^\Lambda(\rho)  &=&  [1 - \xi_{\max}(W)]\;  \sin^2\lambda \;,
\end{eqnarray}
with  $\xi_{\max}(W)$ being the maximum eigenvalue of a $3\times 3$ matrix whose elements are given by
\begin{eqnarray}\label{eq:W}
W_{\alpha\beta} =  \mbox{Tr}[ \sqrt{\rho}\;  \sigma_{A,\alpha}
\sqrt{\rho} \;  \sigma_{A,\beta}]\;.
\end{eqnarray}
If $\rho$ is pure, $\rho=\ketbras{\psi}{\psi}{}$, the discriminating strength reduces to
\begin{eqnarray}\label{eq:qubitquditpure}
D_{A\rightarrow B}^\Lambda(\left| \psi \right\rangle_{ }) = [1 - (q_1 - q_0)^2]\;  \sin^2\lambda \,,
\end{eqnarray}
where $q_1$ and $q_2$ are the Schmidt coefficients of $|\psi\rangle$. In particular, notice that for separable pure states we have $|q_1-q_0| =1$ and the discord vanishes (see property 1 in Sec.~\ref{sec:DS as a measure}). On the other hand, for maximally entangled qubit-qudit states we have $q_0=q_1=1/2$ and the DS reaches the 
maximum value $\sin^2\lambda$ (see property 4).\\

\section{Maximization of the discriminating strength over the set of separable states}\label{sec:sep}
The main role played by the discord in the realm of quantum mechanics is enlightening the presence of those quantum correlations which cannot be classified as quantum entanglement. Here, we investigate the behavior of the discriminating strength when computed on the set of separable states $\rho^{(\mathrm{sep})}$ (yielding zero 
entanglement).
We will prove that for all qubit-qudit systems ($d_A=2$ and $d_B \geq 2$), the maximum discord over the set of separable states  is reached over the subset of  pure {\it Quantum-Classical} (pQC) states given by  convex combinations of pure (non necessarily orthogonal) states $\{ \kets{\psi_k}{A}\}$ on $A$ and orthonormal basis $\{\kets{k}{B}\}$ on $B$, i.e. 
\begin{equation}\label{eq:pQCstates}
\rho^{(\mathrm{pQC})}=\sum_k p_k\;  \ketbras{\psi_{k}}{\psi_{k}}{A}\otimes \ketbras{k}{k}{B}\;, 
\end{equation} 
the $\{ p_k\}$ being probabilities. 
For  the case  $d_B\geq 3$ we have an analytical proof of this fact, which allows us to solve the maximization and show that the following 
identity holds
 \begin{eqnarray}	\max_{\rho^{\mathrm{(sep)}}} D_{A\rightarrow B}^\Lambda(\rho^{\mathrm{(sep)}}) &=& 
	\max_{\rho^{\mathrm{(pQC)}}} D_{A\rightarrow B}^\Lambda (\rho^{\mathrm{(pQC)}}) \nonumber \\
	&=& \frac{2}{3} \sin^2\lambda \;,
 \label{ALLDIM}
\end{eqnarray}
(see Sec.~\ref{ssec:optimal} for the case $d_B=\infty$ and Sec.~\ref{ssec:boundf} for the case $d_M\geq 3$).
For $d_B=2$ (i.e. for the qubit-qubit case) instead  the optimality of the pure-QC states can only be verified numerically showing that 
 \begin{eqnarray}	\max_{\rho^{\mathrm{(sep)}}} D_{A\rightarrow B}^\Lambda(\rho^{\mathrm{(sep)}}) &=& 
	\max_{\rho^{\mathrm{(pQC)}}} D_{A\rightarrow B}^\Lambda (\rho^{\mathrm{(pQC)}}) 
	\nonumber \\&=& \frac{1}{2}	 \sin^2\lambda \;, \label{2DIM}
\end{eqnarray}
(see Sec.~\ref{ssec:qubqub}).

\subsection{pure-QC states maximize the DS over the set of separable states: case $d_B=\infty$}\label{ssec:optimal}
A generic
separable state can always  be written as 
\begin{eqnarray}
\rho^{(\mathrm{sep})} \label{sepstate}
&=& \sum_{k} p_k \ketbras{\psi_{k}}{\psi_{k}}{A} \otimes \rho_B^{(k)}, \end{eqnarray}
where $\{|\psi_k\rangle_A\}$ are (possibly non-orthogonal)
pure states on $\mathcal{H}_A$ and $\{ \rho_B^{(k)}\}$ is a set of density matrices on $\mathcal{H}_B$, while $\{ p_k\}$ are probabilities.
From the joint concavity of the QCB ~(\ref{eq:QCB})~\cite{QCB} and from the cyclic property of the trace, we have
\begin{widetext}
 \begin{eqnarray}
 Q(\rho^{(\mathrm{sep})} ,e^{i H_A^\Lambda} \rho^{(\mathrm{sep})} e^{-i H_A^\Lambda})
 & \geq &\sum_k p_k Q( \ketbras{\psi_{k}}{\psi_{k}}{A}\otimes \rho_B^{(k)} ,e^{i H_A^\Lambda}  \ketbras{\psi_{k}}{\psi_{k}}{A} e^{-i H_A^\Lambda}\otimes \rho_B^{(k)}) \nonumber 
 \\ &=& \sum_k p_k Q(\ketbras{\psi_{k}}{\psi_{k}}{A},e^{i H_A^\Lambda} \ketbras{\psi_{k}}{\psi_{k}}{A}e^{-i H_A^\Lambda})=\sum_k p_k |\bras{\psi_k}{A}e^{i H_A^\Lambda}\ket{\psi_k}_A|^2 \;.
\end{eqnarray}
\end{widetext}
By direct calculation, one can easily verify that the above inequality is saturated a pure-QC state $\rho^{\mathrm{(pQC)}}$ of Eq.~(\ref{eq:pQCstates}) obtained by replacing the 
density matrices $\rho_B^{(k)}$ of~(\ref{sepstate}) with orthogonal projectors $|k\rangle_B\langle k|$
 (notice that this is possible because $B$ is infinite dimensional). 
 Indeed in this case we have 
\begin{widetext}
\begin{align}\label{eq:DSpqc}
Q(\rho^{\mathrm{(pQC)}}, e^{i H_A^\Lambda}& \rho^{\mathrm{(pQC)}} e^{-i H_A^\Lambda})\nonumber \\
& = \min_{0 \leq s \leq 1} \Tr \left[ \left(\sum_k p_k \ketbras{\psi_{k}}{\psi_{k}}{A}\otimes \ketbras{k}{k}{B}\right)^s \left(\sum_{k'} p_{k'} e^{i H_A^\Lambda} \ketbras{\psi_{k'}}{\psi_{k'}}{A}e^{-i H_A^\Lambda}\otimes \ketbras{k'}{k'}{B}\right)^{1-s} \right] \nonumber \\
& = \sum_k p_k |\bras{\psi_k}{A}e^{i H_A^\Lambda}\ket{\psi_k}_A|^2 \;.
\end{align}
\end{widetext}
Since $Q(\rho^{(\mathrm{sep})} ,e^{i H_A^\Lambda} \rho^{(\mathrm{sep})} e^{-i H_A^\Lambda})$ is greater than $Q(\rho^{\mathrm{(pQC)}}, e^{i H_A^\Lambda} \rho^{\mathrm{(pQC)}} e^{-i H_A^\Lambda})$ for each choice of $H_A^\Lambda$, we conclude that
\begin{equation}\label{eq:sep_pQC}
D_{A\rightarrow B}^\Lambda(\rho^{\mathrm{(sep)}})\leq D_{A\rightarrow B}^\Lambda(\rho^{\mathrm{(pQC)}})\,.
\end{equation}
Next we show that 
the maximum DS attainable over the set of pQC states (and hence over the set of separable states) cannot be larger than $\frac{2}{3}\sin^2\lambda$.
To do so let us first consider the \textit{uniform} pQC state $\rho_{u,d}^{\mathrm{(pQC)}}$,
\begin{eqnarray} \label{OPTM}
\rho_{u,d}^{\mathrm{(pQC)}} = \frac{1}{d} \sum_{j=0}^{d-1} \; |\psi_{j}\rangle_A\langle \psi_{j} | 
\otimes |j \rangle \langle j | \;,
\end{eqnarray}
characterized by $d$ pure states $\{ |\psi_j\rangle_A\}$ whose
corresponding vectors $\{ \hat{r}_j\}$ in the Bloch sphere are assumed to be
uniformly distributed (i.e. their $d$ vertices identify
a regular polyhedron). From Eq.~\eqref{eq:DSpqc} we have
 \begin{eqnarray} D_{A\rightarrow B}^\Lambda(\rho_{u,d}^{(\mathrm{pQC})})\!\!\!&=& \!\!\!\min_{\{H_A^\Lambda\}}\sum_{j=0}^{d-1}\frac{1}{d} \left(1- \left | \bras{\psi_j} {A} e^{i H_A^\Lambda}\ket{\psi_j}_A\right|^2\right)\nonumber \\
 &=& \!\!\! \min_{\{\hat{n}\}}\sum_{j=0}^{d-1}\frac{1}{d} \left[1 \!-\! \cos^2\lambda \!- \! \sin^2\lambda \; (\hat{r}_j\cdot\hat{n})^2\right]\nonumber  \\
 &=& \!\!\! \big( 1 -  \max_{\{\hat{n}\}} \frac{1}{d} \sum_{j=0}^{d-1}   \cos^2\theta_j \big)\;\sin^2\lambda \label{eq:pQC_QCB}\;,
\end{eqnarray}
where we set  $H_A^\Lambda=\lambda \sigma_A^{(\hat{n})}$ (see Sec.~\ref{sec:DS_qubit_qudit})  and introduced $\cos\theta_j = \hat{n}\cdot \hat{r}_j$. In the limit \linebreak$d\rightarrow \infty$ the series $\sum_{j=1}^d   \cos^2\theta_j$ converges to an integral over the solid angle, which does not depend on the 
 orientation of $\hat{n}$, i.e. 
 \begin{eqnarray}
\lim_{d\rightarrow \infty}   \frac{1}{d}\sum_{j=0}^{d-1}   \cos^2\theta_j  
&=&
\frac{1}{4\pi} \int d\Omega \cos^2\theta\\ \nonumber 
& =& \frac{1}{4\pi} 
\int_0^{2\pi} \!\!\!\! d\phi \int_0^\pi \!\!\! d\theta \sin \theta \cos^2\theta =\frac{1}{3} \;.
\end{eqnarray}
Therefore we have 
\begin{eqnarray}\label{dueduetrefffd}
	   D_{A\rightarrow B}^\Lambda(\rho_{u,\infty}^{\mathrm{(pQC)}}) = \frac{2}{3}\sin^2\lambda.
\end{eqnarray}
To prove that the above quantity is also the maximum value of DS over
the whole set of pure-QC states~\eqref{eq:pQCstates} we notice that, proceeding as in Eq.~(\ref{eq:pQC_QCB}),  
we can write 
\begin{eqnarray}\nonumber 
  D_{A\rightarrow B}^\Lambda(\rho^{\mathrm{(pQC)}}) \! &=& \!  \big(1-\max_{\{\hat{n}\}}\sum_{j=0}^{d-1} p_j(\hat{r}_j\cdot\hat{n})^2\big)\! \sin^2\lambda   \\
	  &=&  \big(1-\sum_{j=0}^{d-1} p_j(\hat{r}_j\cdot\hat{n}_*)^2\big)\; \sin^2\lambda \;, \label{dueduetrefdfdf2}
 \end{eqnarray}
where $\hat{n}_*$ indicates the direction which is saturating the maximization.
This vector  is clearly a function of the state $\rho^{\mathrm{(pQC)}}$, i.e. it depends on the
probabilities $p_j$ and on the vectors $\hat{r}_j$. If we define the state $\rho_R^{\mathrm{(pQC)}}$, obtained from 
 $\rho^{\mathrm{(pQC)}}$ by applying to the vectors $\hat{r}_j$ 
a rotation matrix $R \in SO(3)$ , we have
\begin{equation}\label{eq:rotatedDS}
D_{A\rightarrow B}^\Lambda(\rho_R^{\mathrm{(pQC)}}) = D_{A\rightarrow B}^\Lambda(\rho^{\mathrm{(pQC)}}),
 \end{equation}
 where the vector saturating the maximization in Eq.~\eqref{dueduetrefdfdf2} now corresponds to $R \hat{n}_*$.
By introducing an ancillary system $C$, associated to the Hilbert space $\mathcal{H}_C$, and a set of $N$ 3D-rotations $\{ R_k \}$, mapping each vertex of the regular N-polyhedron on all vertices (including itself), one can define the density matrix
\begin{eqnarray}\label{expansion1}
&&\bar{\rho}^{\mathrm{(pQC)}}_{ABC}
 \nonumber\\ &&\quad:=\frac{1}{N}\!\sum_{k=0}^{N-1} \!\sum_{j=0}^{d-1} p_j |\psi_{j}(R_k)\rangle_A\langle \psi_{j}(R_k) | \!\otimes\! |j \rangle_B \langle j | \!\otimes\! |k \rangle_C \langle k | \nonumber\\
&&\quad\;=\!\frac{1}{N}\sum_{k=0}^{N-1} \! \rho_{R_k}^{\mathrm{(pQC)}} \!\otimes\! |k \rangle_C \langle k |
 \end{eqnarray}
where
\begin{equation}
\rho_{R_k}^{\mathrm{(pQC)}}:=\sum_{j=0}^{d-1} p_j |\psi_{j}(R_k)\rangle_A\langle \psi_{j}(R_k) | \!\otimes\! |j \rangle_B \langle j |\,.
\end{equation}
On the other hand $\bar{\rho}^{\mathrm{(pQC)}}_{ABC}$ can be also arranged as
   \begin{equation}\label{expansion2}
\bar{\rho}^{\mathrm{(pQC)}}_{ABC}=\!\sum_{j=0}^{d-1} p_j  \rho_{u,N,j}^{\mathrm{(pQC)}} \!\otimes\! |j \rangle_B \langle j |,
 \end{equation} 
 where the density matrices $\rho_{u,N,j}^{\mathrm{(pQC)}}$, on $\mathcal{H}_{A} \otimes \mathcal{H}_{C}$, are defined as
 \begin{equation}
 \rho_{u,N,j}^{\mathrm{(pQC)}}:=\frac{1}{N}\!\sum_{k=0}^{N-1} |\psi_{j}(R_k)\rangle_A\langle \psi_{j}(R_k) |   \!\otimes\! |k \rangle_C \langle k|\,.
 \end{equation}
It is important to observe that since $B$ is infinite dimensional, there always exists a state $\bar{\rho}^{\mathrm{(pQC)}}$ of $AB$ which is fully isomorphic to $\bar{\rho}^{\mathrm{(pQC)}}_{ABC}$, from which it follows
\begin{equation}
Q(\bar{\rho}^{\mathrm{(pQC)}},\hat n) = Q(\bar{\rho}^{\mathrm{(pQC)}}_{ABC},\hat n)\,,
\end{equation}
where
\begin{equation}
Q(\rho,\hat n) := Q\left(\rho, e^{i \lambda \sigma_A^{(\hat{n})}  } \rho e^{-i \lambda \sigma_A^{(\hat{n})}}\right)\,.
\end{equation}
Thanks to expansion \eqref{expansion1}, we get
 \begin{eqnarray}
Q(\bar{\rho}^{\mathrm{(pQC)}},\hat n)=\frac{1}{N} \sum_{k=0}^{N-1} Q({\rho}_{R_k}^{\mathrm{(pQC)}},\hat n),
\end{eqnarray} 
from which, taking the maximum over $\hat n$, it results
\begin{eqnarray}\label{dueffeee}
 \max_{\{\hat n\}} \sum_{k=0}^{N-1} Q({\rho}_{R_k}^{\mathrm{(pQC)}},\hat n) \leq \sum_{k=0}^{N-1} \max_{ \{\hat n\} }Q({\rho}_{R_k}^{\mathrm{(pQC)}},\hat n).
\end{eqnarray}
Finally, since for all $k$, ${\rho}_{R_k}^{\mathrm{(pQC)}}$ and ${\rho}^{\mathrm{(pQC)}}$ share the same DS (see Eq.~\eqref{eq:rotatedDS}), we get
\begin{equation}\label{res1}
D_{A\rightarrow B}^\Lambda(\bar{\rho}^{\mathrm{(pQC)}}) \geq D_{A\rightarrow B}^\Lambda(\rho^{\mathrm{(pQC)}}).
\end{equation}
On the other hand, thanks to expansion \eqref{expansion2} we have
\begin{eqnarray}
Q(\bar{\rho}^{\mathrm{(pQC)}},\hat n) =\sum_{j=0}^{d-1} p_j Q \left( \rho_{u,N,j}^{\mathrm{(pQC)}},\hat n \right)\,.
\end{eqnarray} 
and therefore
 \begin{eqnarray}\label{dueffeee2}
 \max_{\{\hat n\}} Q(\bar{\rho}^{\mathrm{(pQC)}},\hat n) \leq \sum_{j=0}^{d-1} p_j \max_{\{\hat n\}} Q\!\! \left( \rho_{u,N,j}^{\mathrm{(pQC)}},\hat n \right)\!.
\end{eqnarray}
The above inequality is saturated in the limit $N \rightarrow \infty$, where each $\rho_{u,N,j}^{\mathrm{(pQC)}}$ approaches the state $\rho_{u,\infty}^{\mathrm{(pQC)}}$ characterized by
\begin{equation}
Q \big(\rho_{u,\infty}^{\mathrm{(pQC)}}, \hat n\big) = \cos^2\lambda  - \frac{1}{3}\sin^2\lambda, \quad \forall \hat{n}\,
\end{equation}
(see Eq.~\ref{dueduetrefffd}).
We therefore have
\begin{equation}\label{res2}
D_{A\rightarrow B}^\Lambda(\bar{\rho}^{\mathrm{(pQC)}}) \stackrel{N \rightarrow \infty}{=} \sum_{j=0}^{d-1} p_j \frac{2}{3}\sin^2\lambda= \frac{2}{3}\sin^2\lambda.
\end{equation}
The identity~(\ref{ALLDIM}) finally follows by
combining Eqs.~\eqref{eq:sep_pQC}, \eqref{res1} and \eqref{res2}.

\subsection{pure-QC states maximize the DS over the set of separable states: case $d_B\geq 3$}\label{ssec:boundf}

If $\mathcal{H}_B$ is finite dimensional we are not guaranteed about the possibility of mapping a generic separable state in the a pure-QC state. Thus relation~\eqref{eq:sep_pQC} could be in principle violated. However by embedding $\mathcal{H}_B$ into a larger system having infinite dimension one can still  invoke the result of the previous subsection to say that  
  \begin{eqnarray}	\max_{\rho^{\mathrm{(sep)}}} D_{A\rightarrow B}^\Lambda(\rho^{\mathrm{(sep)}})
\leq \frac{2}{3} \sin^2\lambda \;.
 \label{ALLDIMineq}
\end{eqnarray}
To prove Eq.~(\ref{ALLDIM}) it is hence sufficient to produce an example of a pure-QC state   (\ref{eq:pQCstates}) 
that achieves such upper bound. Of course  the sequence of uniform states (\ref{OPTM})
cannot be used for this purpose because now $d_B$ is explicitly assumed to be finite. Instead 
we take 
\begin{eqnarray} \label{ttff}
\rho^{(p\mathrm{QC})} = \sum_{j=0,1,2} p_j\;  |\psi_j\rangle_A \langle \psi_j|
\otimes |j\rangle_B\langle j|\;,
\end{eqnarray} 
with $|0\rangle_B$, $|1\rangle_B$, $|2\rangle_B$ being orthonormal  elements of $\mathcal{H}_B$, 
which is a properly defined p-QC state whenever  the dimension $d_B$ is larger than 3. 
As in the first line of Eq.~(\ref{dueduetrefdfdf2}),  its associated 
discriminating strength  can be  then computed as, 
 \begin{eqnarray}\label{eq:dueduetre}
 D_{A\rightarrow B}^\Lambda(\rho^{(\mathrm{pQC})})
=  \big(1-\max_{\hat{n}}\sum_{j=0}^2 p_j(\hat{r}_j\cdot\hat{n})^2\big)\; \sin^2\lambda  \,,
\end{eqnarray}
where $\hat{r}_j$ is the vector in the Bloch sphere of the state $\ket{\psi_j}$ while $H_A^\Lambda=\lambda \sigma_A^{(\hat{n})}$. 
We are interested in the case where   $\{\hat{r_j}\}$
is an orthonormal triplet (i.e. the three vectors identifying three Cartesian axes in the 3D-space). Notice that this does not mean that the corresponding states are orthogonal: instead they are mutually
unbalanced states (e.g.  $|\psi_0\rangle_A=|0\rangle_A$, 
 $|\psi_1\rangle_A=|+\rangle_A=(\ket{0}_A + \ket{1}_A)/\sqrt{2}$,  $|\psi_2\rangle_A=|\times \rangle_A = (|0\rangle_A + i |1\rangle_A)/\sqrt{2}$), so that (\ref{ttff})  corresponds to an (unbalanced) Generalized B92 (GB92) state~\cite{B92}. From the normalization condition on vector $\hat{n}$, it derives that the squared scalar products $(\hat{n}\cdot \hat{r}_j)^2$
 define a set of probabilities, since
 \begin{eqnarray}
 \sum_{j=0,1,2} (\hat{n}\cdot \hat{r}_j)^2  = |\hat{n}|^2 =1
 \;.
 \end{eqnarray} 
Thus, the maximization involved in (\ref{eq:dueduetre}) can be trivially performed by choosing $\hat n$ parallel to the $\hat r_j$ associated to the  maximum weight $p_j$. This gives
  \begin{eqnarray} 
  D_{A\rightarrow B}^\Lambda(\rho^{(\mathrm{GB92})}) =  \left(1 - \max\{ p_0, p_1,p_2\}\right)\; \sin^2 \lambda\;.
 \end{eqnarray} 
By observing that for a three event process the maximum probability can never be smaller than $1/3$, we conclude that  the maximum DS over the set of GB92 states is achieved by the Equally Weighted (EW) one
\begin{eqnarray}\label{eq:GB92sat}
\rho_{\mathrm{EW}}^{(\mathrm{GB92})} &=&\frac{1}{3} \Big( \ketbras{0}{0}{A}  \otimes \ketbras{0}{0}{B}+ \ketbras{+}{+}{A}  \otimes \ketbras{1}{1}{B}\nonumber \\
&& \quad+  \ketbras{\times}{\times}{A}  \otimes \ketbras{2}{2}{B} \Big)\;.
\end{eqnarray}
With this choice we get 
\begin{eqnarray} 
    D_{A\rightarrow B}^\Lambda(\rho_{\mathrm{EW}}^{(\mathrm{GB92})})=\frac{2}{3} \sin^2 \lambda \;, \label{GB92}
 \end{eqnarray} 
 which shows that, also for $d_B$ finite and larger than 3,  the upper bound~(\ref{ALLDIMineq}) is achievable with a pure-QC state, hence proving~(\ref{ALLDIM}).

\subsection{p-QC states maximize the DS over the set of separable states: case $d_B=2$ (qubit-qubit) }
\label{ssec:qubqub}

The argument used in the previous section cannot be directly applied to analyze the qubit-qubit case  (i.e. $d_A=d_B=2$), because for those systems the states~(\ref{ttff}) 
and (\ref{eq:GB92sat}) cannot be defined. Furthermore we will shall see that the upper bound~(\ref{ALLDIMineq}) is no longer tight.  
To deal with this case we first consider the class of QC state and show the maximum of DS,  equal to $(1/2) \sin^2\lambda$, is achieved on the set of pure-QC states. Then we 
resort to numerical optimization procedures to show that no other separable qubit-qubit state can do better than this, hence verifying the identity~(\ref{2DIM}).  

\subsubsection{Maximum DS over QC states}
A generic QC state for the qubit-qubit case can be expressed as 
\begin{equation}
\rho^{(\mathrm{QC})} \!=\!  p \;  \tau_0 \!\otimes\! |0\rangle_B\langle 0| + (1-p)  \;  \tau_1  \!\otimes\! |1\rangle_B\langle 1|,
\end{equation}
where $p\in [0,1]$, $\tau_{0}$ and $\tau_1$ are generic mixed state of $A$, and  $\{ \ket{0}_B,\ket{1}_{B}\}$ is an orthonormal basis of $\mathcal{H}_B$. 
To compute the associated value of  DS we invoke Eq.~(\ref{EQNUOVA}) and determine  
the maximum eigenvalue of the matrix
$W_{\alpha\beta}$ of Eq.~\eqref{eq:DS_LQU}.
Recalling the invariance of DS under local unitary operations we then set 
\begin{equation}
\tau_0 \!=\! \frac{I + s_0 \sigma_3}{2}, \quad \tau_1 \!=\! \frac{I + s_1 ( \sin\phi \; \sigma_1 +  \cos\phi \;  \sigma_3)}{2},
\end{equation}
with $0\leq \phi \leq \pi$ and $0\leq s_i \leq 1$, which yields
\begin{equation}
\sqrt{\tau_i} = R(\phi_i)\frac{A(s_i) + B(s_i)\sigma_3}{\sqrt{2}} R^\dagger(\phi_i),
\end{equation}
where
$\phi_0=0$, $\phi_1=\phi$, $R(\theta)=\exp\left[-i\frac{\theta}{2} \sigma_2\right]$ and
\begin{equation}
A(s_i)\!=\!\frac{\sqrt{1+s_i} + \sqrt{1-s_i}}{2}, \; \;B(s_i)\!=\!\frac{\sqrt{1+s_i} - \sqrt{1-s_i}}{2}.
\end{equation}
We now have all the ingredients necessary for the computation of the matrix elements $W_{\alpha\beta}$. Thanks to the orthogonality of $|0\rangle_B$ and $|1\rangle_B$,
this gives 
\begin{widetext}
\begin{eqnarray}
W_{2\beta}
&=& W_{\beta2} = \left[p\sqrt{1-s_0^2}+(1-p)\sqrt{1-s_1^2}\right]\delta_{2\beta}\, >0 \nonumber\\
W_{11}&=&p\sqrt{1-s_0^2} + \frac{(1-p)}{2}\left[ 1-\cos(2\phi)  + \sqrt{1-s_1^2}\,(1+\cos(2\phi)) \right] >0\nonumber \\
W_{13}&=&W_{31} = (1-p)\left(1-\sqrt{1-s_1^2}\right)\sin\phi \;\cos\phi \nonumber \\
W_{33}&=&\frac{1+p}{2}+ \frac{1-p}{2}\left[\cos(2\phi)+  \sqrt{1-s_1^2} \; (1-\cos(2\phi))\right].\nonumber\\
\end{eqnarray}
\end{widetext}
It derives that the eigenvalues of $W$ reduce to
\begin{eqnarray}
\xi_0 &=& W_{22} \nonumber\\
\xi_{\pm} &=& \frac{W_{11}+W_{33}}{2} \pm \frac{1}{2}\sqrt{\left(W_{11}-W_{33}\right)^2 + 4 W_{13}^2}. \nonumber
\end{eqnarray}
Being $W_{22}<1$ and $W_{11}+W_{33}=1+W_{22}$, we have that
$\xi_+$ is the maximum eigenvalue. Therefore Eq.~\eqref{EQNUOVA} yields 
\begin{eqnarray}
D_{A\rightarrow B}^\Lambda(\rho^{(\mathrm{QC})} )=  f_W \;\frac{\sin^2\lambda}{2}\;,
\end{eqnarray}
where
\begin{equation}
f_W:= 1- W_{22}-\sqrt{\left(W_{11}-W_{33}\right)^2 + 4 W_{13}^2}\,.
\end{equation}
It  derives 
\begin{eqnarray}
D_{A\rightarrow B}^\Lambda(\rho^{(\mathrm{QC})} )\leq \;\frac{\sin^2\lambda}{2}\,,
\end{eqnarray}
the equality being saturated when $W_{22}=0$, $W_{13}=0$ and $W_{11}-W_{33}=0$. The first condition sets to $1$ the purity of $\tau_0$ and $\tau_1$ ($s_0^2=s_1^2=1$), the second and third conditions imply $\phi=(2n+1)\pi/2$, with $n \in \mathbb{Z}$, and $p=1/2$.
We conclude that the maximum of the DS on the set of QC states is achieved on B92-like states, which are pure-QC, that is
\begin{equation}
\max_{\rho^{\mathrm{(QC)}}} D_{A\rightarrow B}^\Lambda( \rho^{\mathrm{(QC)}}) = D_{A\rightarrow B}^\Lambda(\rho^{\mathrm{(B92)}})=\frac{\sin^2\lambda }{2}\,,
\end{equation}
being
\begin{equation}\label{eq:b92state}
\rho^{{(\mathrm{B92})}}\!=\! \frac{1}{2} \left( |0\rangle_A \langle 0| \!\otimes\! |0\rangle_B
\langle 0|+ |\sin({\phi})\rangle_A\langle \sin({\phi})| \!\otimes\! |1\rangle_B 
\langle 1|\right)\,,
\end{equation}
and $\sin({\phi})=\pm1$ and $\ket{\pm}=(\ket{0}\pm \ket{1})/2$.

\subsubsection{Separable qubit-qubit states: numerical results}\label{sec:num_qubitqubit}
We conclude our analysis by providing numerical evidence that $(1/2) \sin^2\lambda$ is the maximum value reached by the discriminating strength over the all set of separable states as
anticipated in Eq.~(\ref{2DIM}). 
We recall that a generic separable state of two qubit systems can always be written as a finite convex sum of direct products of pure states for $A$ and $B$ \cite{Sanpera}, i.e. 
\begin{eqnarray}\label{eq:sep}
\rho^{(\mathrm{sep})} \!=\! \sum_{j=1}^{N} p_j  \ketbras{\psi_j}{\psi_j}{A}  \otimes \ketbras{\chi_j}{\chi_j}{B}, \quad p_j > 0 \;\; \forall j,
\end{eqnarray}
with $1\leq N \leq 4$. We remark that here no orthogonality constraint has to be imposed on either sets of pure states $\{\left|\psi_j\right>_A\}$ and $\{\left|\chi_j\right>_B\}$, on $\mathcal{H}_A$ and $\mathcal{H}_B$, respectively. The Bloch sphere formalism allows us to define, for all $j$
\begin{eqnarray}
\ketbras{\psi_j}{\psi_j}{A}\!:=\! \frac{\mathbb{I} + \hat{u}_j \cdot \vec{\sigma}_A}{2} \quad \!\mbox{and}\! \quad \ketbras{\chi_j}{\chi_j}{B}\!:=\!\frac{\mathbb{I} + \hat{v}_j \cdot \vec{\sigma}_A}{2}\,. \nonumber\\
\end{eqnarray}
Summarizing, all qubit-quibit separable states are characterized by a set of $N$ probabilities and $2 N$ vectors of unit norms. 

\begin{figure}[ht]
	\includegraphics[width=0.48\textwidth]{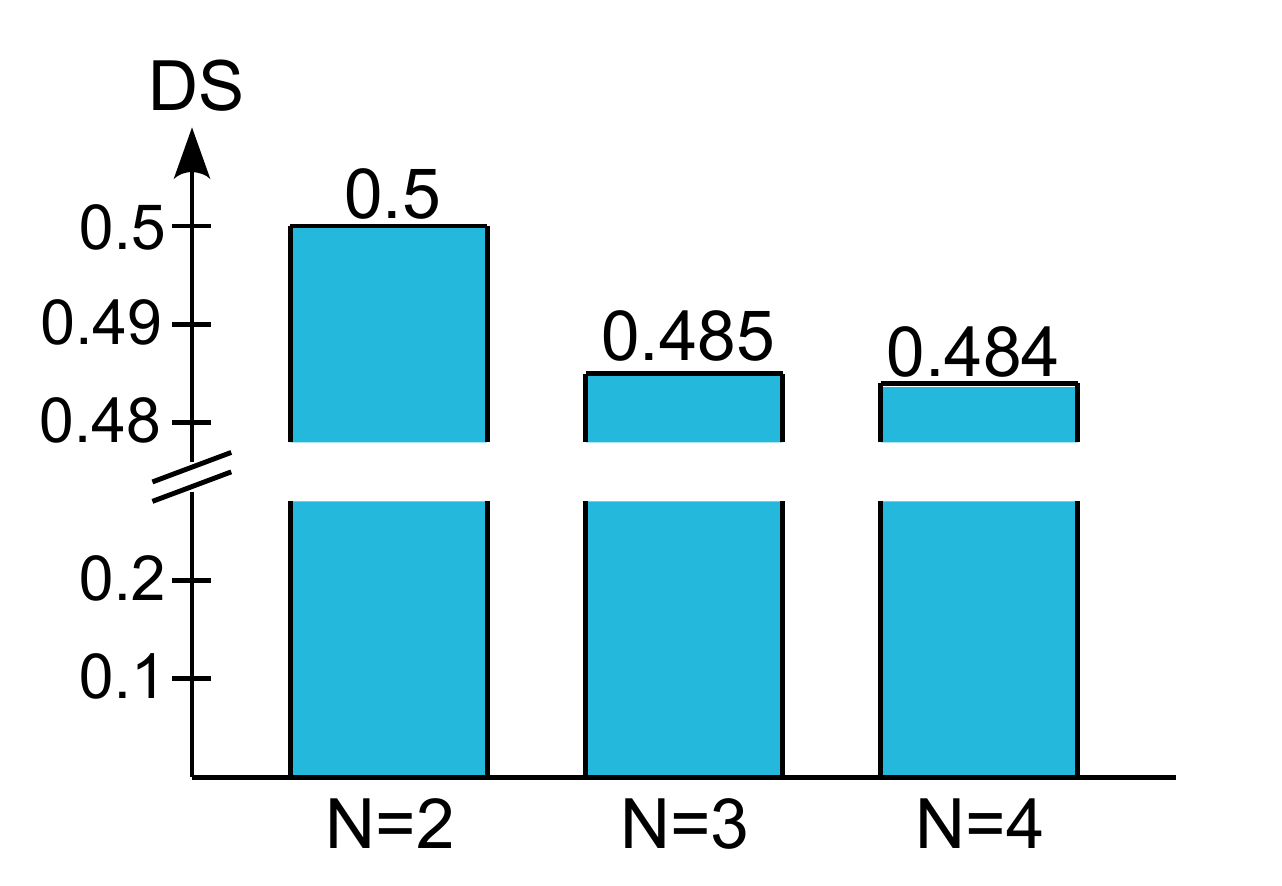}
		\caption{Histogram of the data referring to the numerical computation of ${D_{A\rightarrow B}^\Lambda(\rho^{(\mathrm{sep})}})/\sin^2(\lambda\varphi)$ for qubit-qubit separable states, corresponding to $N=2,3,4$  in Eq.~\eqref{eq:sep}. }
		\label{fig:data}
\end{figure}

The case $N=1$ is trivial (all separable states are completely uncorrelated) and the DS is always zero. Therefore, we have numerically analyzed the cases $N=2$, $N=3$ and $N=4$ and plot our results in Fig.~\ref{fig:data}. The reported results are in agreement with Eq.~(\ref{2DIM}). 

The details of this numerical analysis are presented in Appendix~\ref{app:numerics}.

\section{Conclusions}\label{Sec:conc}
In this  paper we have introduced, under the name of \textit{discriminating strength}, a novel measure of discord-like correlations, i.e. correlations that, even though not being addressable as quantum entanglement, are still non-classical. In the \textit{mare-magnum} of definitions and measures ~\cite{MODIREV}, each stemming from a different way in which quantum correlations can be used to outperform purely classical systems, the discriminating strength finds its natural collocation in the context of state discrimination. More precisely, it quantifies the ability of a given bipartite probing state to discriminate between the application or not of a unitary map to one of its two subsystems, when a large number of copies of the probing state is at disposal. We report that in a similar context, the noisy quantum illumination~\cite{TAN}, a recent paper~\cite{WEED} has put forward a connection between the advantage yielded by quantum illumination over the best conceivable classical approach, and the amount of quantum discord (as in Ollivier and Zurek ~\cite{OLLZU}) surviving in a maximally entangled state after the interaction with a noisy environment. Here however, our goal was to define a quantity which has a clear operative meaning (characterizing quantitatively each bipartite state as a resource for a specific task) and is also easy to compute, at least in some simple cases.

Specifically, we have proved that the discriminating strength fits all the requirements ascribing it as a proper measure of quantum correlations ~\cite{MODIREV}. We have also provided a closed expression of this measure for some special cases, such as pure states and qubit-qudit systems. For the latter case we have also shown an explicit connection with another measure of quantum correlations, the local quantum uncertainty \cite{LQU}, which, in the most general case, can be seen to approximate the discriminating strength in the limit where the unitary map is close to the identity. Next, we have focused on the class of separable states and proved, by means of both analytical and numerical methods, that for all qubit-qudit systems the discriminating strength reaches its maximum on the set of pure quantum classical states. Finally, we have explicitly determined this maximum value.

We remind that by definition the discriminating strength depends on the spectral properties of the encoding Hamiltonian $H_A^\Lambda$. In other words, for each specific choice of $\Lambda$ one can in principle define a different measure of quantum correlations (a similar problem also affects the local quantum uncertainty). It would be therefore interesting to investigate if there exists a criterion for comparing different measures arising from different spectral properties of $H_A^\Lambda$.

To conclude, we remark that the discriminating strength can be related to other discord-like measures that have been recently introduced, including the inteferometric power~\cite{blindmetr}, the local quantum uncertainty~\cite{LQU} and the discord of response~\cite{discResp}. Ultimately, all these measures share a common message: discord-like correlations are the fundamental resource to be used in many quantum metrology tasks. Moreover, the functionals on which they are based (Chernoff bound, Fisher information, Bures distance) are all interconnected, so that each measure could be used to bound the others~\cite{QCB,distances,QCBqubitAndGauss}. Most interestingly, even the Bures geometric quantum discord, which stems from a different perspective, has been recently shown to be related to an ambiguous state discrimination problem~\cite{Spehner2013a}. In this perspective, we believe that our analysis marks a further step towards a novel classification  of a vast set of non-classicality measures.

\acknowledgements 
We thank G. Adesso, D. Girolami, F. Illuminati and T. Tufarelli for useful comments and discussions. 
ADP acknowledges support from Progetto Giovani Ricercatori 2013 of SNS.

\appendix
\section{Pedagogical remark}\label{AppA}
In this appendix, we provide an explicit proof that an arbitrary qubit-qutrit pQC state~\eqref{ttff} cannot achieve a DS greater than $(2/3)\sin^2\lambda$. Note that this result naturally derives from what found in Secs.~\ref{ssec:optimal} and \ref{ssec:boundf}. Nonetheless, we report the following proof as a pedagogical remark for the interested reader.

\begin{figure}[htbp]
	\includegraphics[trim=0pt 0pt 0pt 0pt, clip, width=0.2\textwidth]{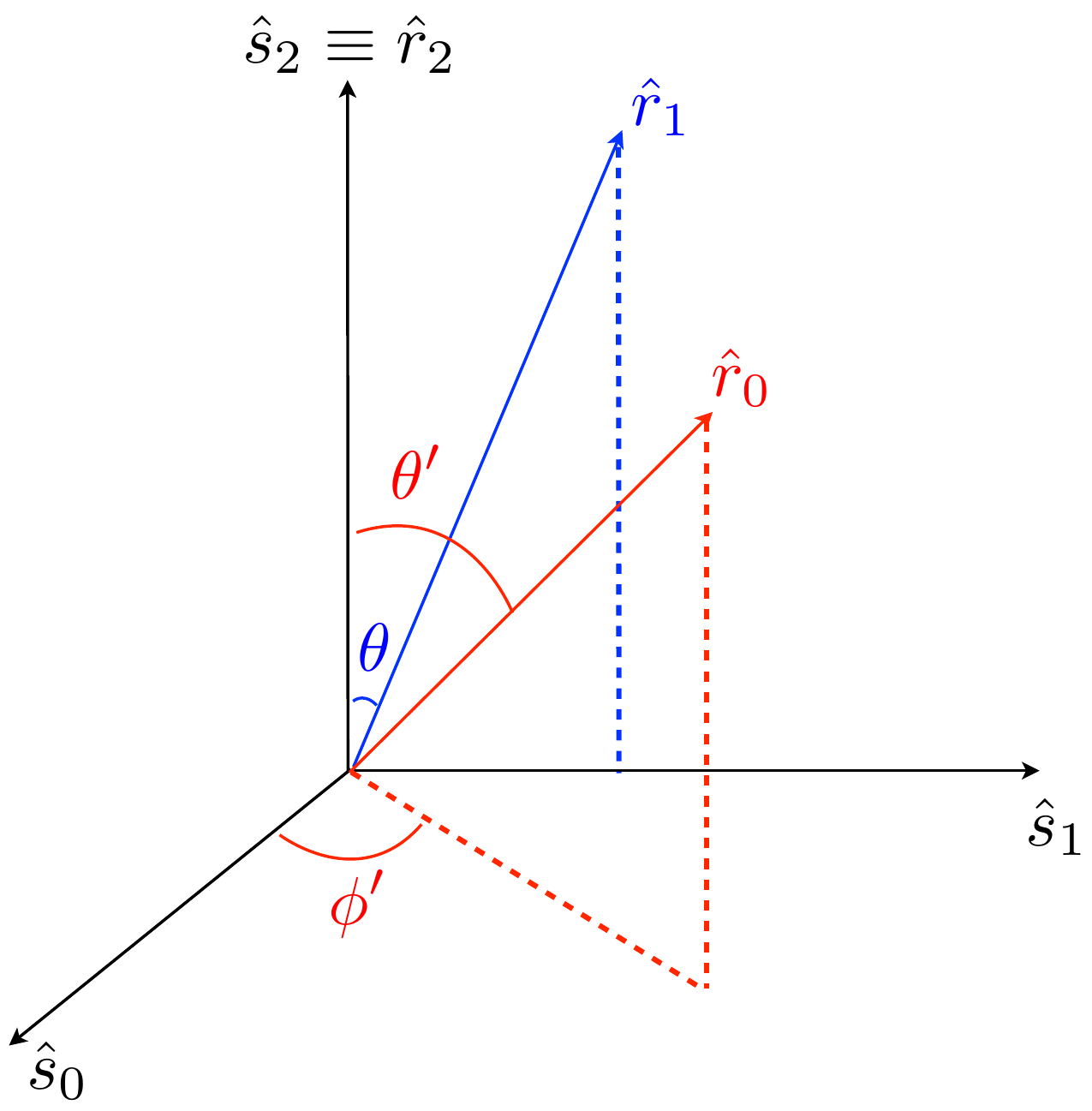}
		\caption{Bloch sphere representation of the qubit pure states $\{\left| \psi_j \right>_A\}$ whit associated unit vectors $\{\hat{r}_j\}$. A Cartesian reference frame $\{\hat{s}_j\}$ is also shown.}
		\label{s_space}
\end{figure}

Consider an arbitrary qubit-qutrit pQC state~(\ref{ttff}) with 
strictly positive  probabilities  $\{ p_j\}$ and with vectors $\{\hat{r}_j\}$ lying in the Bloch sphere. Without loss of generality we assume that $p_2\geq p_1 \geq p_0$ and introduce a Cartesian coordinate set formed by the 3D orthonormal vectors
$\{\hat{s}_j\}$ such that 
\begin{eqnarray}
\hat{r}_2&=& \hat{s}_2\;, \nonumber \\
\hat{r}_1 &=& \cos \theta \hat{s}_2  +\sin\theta \hat {s}_1 \;. \nonumber \\
\hat{r}_0 &=& \cos \theta' \hat{s}_2  +\sin\theta' \cos\phi'  \hat {s}_0 + 
\sin\theta' \sin\phi' \hat{s}_1\;,
\end{eqnarray} 
See Fig.~\ref{s_space}.
With this choice we can write 
\begin{eqnarray} \label{rre}
 \sum_{j=0,1,2}  \! p_j (\hat{n}\cdot \hat{r}_j)^2 = \!\!\! \sum_{j=0,1,2}  \! \tilde{p}_j  \cos^2\phi_j  +
  \Delta(\phi_0,\phi_1,\phi_2)  \;,
 \end{eqnarray} 
 where $\phi_j$ is the angle between $\hat{n}$
and the Cartesian $j$-th axis $\hat{s}_j$,  
\begin{eqnarray} \label{coord}
\cos\phi_j  = \hat{n} \cdot \hat{s}_j \;,
\end{eqnarray}  
$\{\tilde{p}_j\}$ is still a probability set of elements 
 \begin{eqnarray} 
 \tilde{p}_2 &=& p_2 + p_1 \; \cos^2\theta + p_0 \; \cos^2\theta' \;,\nonumber \\
\tilde{p}_1 &=& p_1 \; \sin^2\theta + p_0 \; \sin^2\theta' \sin^2 \phi' \;, \nonumber \\
\tilde{p}_0  &=&  p_0 \; \sin^2\theta'  \cos^2\phi'\;, \nonumber \\
 \end{eqnarray} 
 and $\Delta(\phi_0,\phi_1,\phi_2)$ is the function
 \begin{eqnarray}
 \Delta(\phi_0,\phi_1,\phi_2) \!\!&=& \!\!A \cos \phi_2 \cos\phi_1 + B \cos\phi_2 \cos\phi_0 \nonumber \\&+& \!\!C \cos \phi_0 \cos\phi_1\;, \label{defDELTA} \\
 \nonumber \\
 A&=& p_1 \sin2 \theta + p_0 \sin2\theta'\sin\phi' \nonumber \\ 
B&=& p_0 \sin 2\theta' \cos\phi' \nonumber \\
C&=& p_0 \sin^2\theta' \sin2\phi'  \;.
 \end{eqnarray} 
 Observe that all the dependence of~(\ref{rre})
   upon $\hat{n}$ relies on the phases $\{\phi_j\}$: in particular
the probabilities  $\{ \tilde{p}_j\}$ and the quantity $A$, $B$, and $C$
 of Eq.~(\ref{defDELTA}) do not depend on the choice of the Hamiltonian:
 they only depend on the initial state~(\ref{ttff}). 
 
 According to~\eqref{eq:dueduetre}, in order to compute the discriminating strength of the state we need to find the maximum value of~(\ref{rre}) over all 
possible choices of $\hat{n}$, i.e. for all possible coordinates components~(\ref{coord}). 
To do so we first  use the following facts to show that 
it is always possible to have $\Delta$ positive while
keeping the first contribution of~(\ref{rre}) positive 
(i.e. $\sum_{j=0,1,2}  \tilde{p}_j  \cos^2\phi_j\geq0$):
\begin{itemize}
\item[{\bf F1:}] given  three real number $a$, $b$ and $c$,
at least one of the four combination must be non negative, 
i.e. $a+b+c$, $a-b-c$, $-a+b-c$, $-a-b+c$ (observe that their 
sum is null);
\item[{\bf F2:}] The  vectors which with respect to $\{ \hat{s}_j\}$ have coordinates 
\begin{eqnarray} 
 \hat{n}_1&:=& (\cos\phi_0,\cos\phi_1,\cos\phi_2)\;,\nonumber \\
\hat{n}_2&:=&(-\cos\phi_0,\cos\phi_1,\cos\phi_2)\;,\nonumber \\
\hat{n}_3&:=&(\cos\phi_0,-\cos\phi_1,\cos\phi_2)\;, \nonumber \\
\hat{n}_4&:=&(\cos\phi_0,\cos\phi_1,-\cos\phi_2)\;,\nonumber 
\end{eqnarray}
have the same value of $\sum_{j=0,1,2}  \tilde{p}_j  \cos^2\phi_j$ but 
are associated to the following values for
 $\Delta(\phi_1,\phi_2,\phi_3)$,
\begin{eqnarray}
\hat{n}_1& \mapsto& \Delta= a+b+c\;,\nonumber \\
\hat{n}_2& \mapsto&  \Delta= a-b-c\;,\nonumber \\
\hat{n}_3& \mapsto&  \Delta=-a+b-c\;,\nonumber \\
\hat{n}_4& \mapsto&  \Delta=-a-b+c\;,
\end{eqnarray} 
with $a=A\cos \phi_2 \cos\phi_1$,
$b=B \cos\phi_2 \cos\phi_0$ and $c=C \cos \phi_0\cos\phi_1$.
From {\bf F1} it derives that at least one of the vectors $\hat{n}_{1,2,3,4}$
will have $\Delta$ positive.
\end{itemize}
We therefore conclude that 
\begin{eqnarray} \label{rre11}
\max_{\hat{n}}  \sum_{j=0,1,2}  p_j  (\hat{n}\cdot \hat{r}_j)^2  &\geq& \max_{\hat{n}}  \sum_{j=0,1,2}  \tilde{p}_j  \cos^2\phi_j  \nonumber \\
& =&
\max \{ \tilde{p}_0,   \tilde{p}_1,   \tilde{p}_2\}\;,
 \end{eqnarray} 
where the last identity follows from the fact that 
$\{\cos^2\phi_j\}$ is a probability set, since it 
fulfills the normalization condition~$\sum_{j=0,1,2} \cos^2\phi_j=1$, see Eq.~(\ref{coord}).
Replacing this into Eq.~\eqref{eq:dueduetre} finally yields
\begin{eqnarray}\label{dueduetretre}
	  D_{A\rightarrow B}^\Lambda(\rho^{(\mathrm{pQC})})  \!\! &\leq& \!\!  (1 - \max \{ \tilde{p}_0,   \tilde{p}_1,   \tilde{p}_2\} ) \sin^2\lambda \nonumber \\ 
	  &\leq& \!\! \frac{2}{3}\sin^2\lambda \;, 
\end{eqnarray}
where the last inequality holds because the largest of three positive quantities summing to $1$ cannot be smaller than $1/3$.

\section{Numerical analysis for qubit-qubit separable states}\label{app:numerics}
This appendix is devoted to discussing in deeper details the numerical analysis presented in Sec.~\ref{sec:num_qubitqubit}.

We have computed the discriminating strength of a two-qubit system in an arbitrary separable state, which, without loss of generality can be written as
\begin{eqnarray}\label{eq:22SEPBloch}
\rho^{\mathrm{(sep)}} \!=\! \sum_{j=1}^{N} p_j \frac{\mathbb{I} + \hat{u}_j \cdot \vec{\sigma}_A}{2} \otimes \frac{\mathbb{I} + \hat{v}_j \cdot \vec{\sigma}_B}{2}, \quad p_j > 0 \;\; \forall j\,, \nonumber\\
\end{eqnarray}
with $1\leq N \leq 4$, and $\hat{u}_j,\hat{v}_j$ normalized vectors in the Bloch sphere~\cite{Sanpera}.

Let us start with the case $N=2$.  The set of probabilities $\{p_i\}$ can be labelled as
\begin{eqnarray}
 \{p_1,p_2\}&=&C_2\{\sin\alpha,\cos\alpha\}\nonumber\\
 C_2&=&\frac{1}{\sin\alpha+\cos\alpha}\,,
\end{eqnarray}
with $0<\alpha\leq \pi/4$. The latter constraint implies $0<p_1\leq p_2$. Similarly, we have parametrized the unit vectors $\hat{u}_j$ and $\hat{v}_j$ by means of the polar and  azimuthal angles, $0 \leq \theta^{u,v}_j \leq \pi$ and $0 \leq \phi^{u,v}_j < 2\pi$, respectively. For each angle, we have taken a set of uniformly distributed values within the corresponding range, and perform all possible combinations. Finally, we have set some additional constraints in the numerical code in order get rid of those states which are equivalent under local unitary transformations. 
Thanks to this procedure, we have generated a set of $\sim 7 \times 10^8$ separable states and found that the state with maximum DS corresponds to the B92 state \eqref{eq:b92state} with  $D_{A\rightarrow B}^\Lambda=1/2 \sin^2(\lambda\varphi)$, thus confirming what shown in Sec.~\ref{ssec:qubqub}.

We have repeated the same analysis for the case $N=3$ by setting  
\begin{eqnarray}
\{p1,p2,p3\}&=&C_3\{\sin\alpha\sin\beta,\sin\alpha\cos\beta,\cos\alpha\} \nonumber\\
 C_3&=&\frac{1}{\sin\alpha(\sin\beta+\cos\beta)+\cos\alpha}
\end{eqnarray}
with $0<\alpha,\beta \leq \pi/4$ to ensure that $0<p_1\leq p_2\leq p_3$. 
We thus generated a set of $\sim 2 \times 10^6$ separable states. The maximum DS detected within this ensemble is $\sim 0.485 \sin^2(\lambda\varphi)$, and corresponds to
\begin{eqnarray}
&& \alpha=3\pi/16, \beta=\pi/4, \nonumber \\
&&\theta^{u,v}_j=\phi^{u,v}_j=0, \; \mbox{for} \; j=1,2 \nonumber \\
 &&\theta^{u}_3=\phi^{u}_3=\pi/2, \; \theta^{v}_3=\pi, \phi^{v}_3=0\,.
 \end{eqnarray}
Up to local unitary transformations, this set of parameters describes the state
\begin{eqnarray}
\rho^{\mathrm{(sep)}}\simeq0.486 \ketbras{0}{0}{A} \otimes \ketbras{0}{0}{B} + 0.514 \ketbras{+}{+}{A} \otimes \ketbras{1}{1}{B},\nonumber\\
\end{eqnarray}
which is almost equivalent the B92 state~\eqref{eq:b92state} found for $N=2$.
We foresee that, by means of a finer graining of the parameter space, one should be able to include in the ensemble generated with this procedure the B92 state and reach $1/2 \sin^2(\lambda\varphi)$ as the highest value for DS.\\

Finally we considered the case $N=4$, which corresponds to setting in Eq.~\eqref{eq:22SEPBloch}
\begin{eqnarray}
\{p1,p2,p3,p4\}&=&C_4\{\sin\alpha\sin\beta\sin\gamma,\sin\alpha\sin\beta\cos\gamma,
\nonumber\\ &&\qquad\qquad\sin\alpha\cos\beta,\cos\alpha\} \nonumber\\ \nonumber\\
C_4=\sin\alpha(\sin\beta\!\!\!\!&(&\!\!\!\!\sin\gamma+\cos\gamma)+\cos\beta) +\cos\alpha
\end{eqnarray}
with $0<\alpha,\beta,\gamma \leq \pi/4$ ensuring $0<p_1\leq p_2\leq p_3 \leq p_4$. We have thus generated a set of $\sim 10^6$ separable states. The maximum value we have found for the discriminating strength  is $\sim 0.484 \sin^2(\lambda\varphi)$, achieved when
\begin{eqnarray}
&&\alpha=\pi/4, \; \beta=\pi/8, \; \gamma=\pi/4 \nonumber\\
&&\theta^{u,v}_j=0, \phi^{u,v}_j=0,  \quad \mbox{for} \; j=1,4 \nonumber \\
&&\theta^{u}_k=\pi/2, \; \theta^{v}_k=\pi, \; \phi^{u,v}_k=0, \; \mbox{for} \;  k=2,3\,.\nonumber\\
\end{eqnarray}
This set of parameters defines the state
\begin{eqnarray}
\rho^{\mathrm{(sep)}}& \simeq 0.515\ketbras{0}{0}{A} \otimes \ketbras{0}{0}{B}  +  0.485 \ketbras{+}{+}{A} \otimes \ketbras{1}{1}{B}\,,\nonumber\\
\end{eqnarray} 
which again, up to numerical errors, is quite close to the aforementioned B92 state.

\end{document}